\documentstyle[preprint,aps,prb,epsf,eqsecnum]{revtex}

\tighten

\begin{document}

\draft

\title{The orbital magnetization of single and double quantum dots in a
tight binding model}

\author{A. Aldea$^1$, V. Moldoveanu$^1$, M. Ni\c t\u a$^1$, 
A. Manolescu$^{1,2}$, V. Gudmundsson$^2$, B. Tanatar$^3$}

\address{
$^1$ National Institute of Materials Physics, POBox MG7,
Bucharest-Magurele, Romania \\
$^2$ Science Institute, University of Iceland, Dunhaga 3, IS-107 
Reykjavik, Iceland \\
$^3$ Department of Physics, Bilkent University, 06533 Ankara, Turkey}

\maketitle

\begin{abstract}
We calculate the orbital magnetization of single and double quantum
dots coupled both by Coulomb interaction and by electron tunneling.
The electronic states of the quantum dots are calculated in a 
tight-binding model and the magnetization is discussed in relation to 
the energy spectrum and to the edge and bulk states.  We identify 
effects of chirality of the electronic orbits and of the anti-crossing 
of the energy levels when the magnetic field is varied.  We also 
consider the effects of detuning the energy spectra of the quantum
dots by an external gate potential. We compare our results with the 
recent experiments of Oosterkamp {\it et al}.
[Phys. Rev. Lett. {\bf 80}, 4951 (1998)].
\end{abstract}


\section{Introduction}

The magnetic properties of mesoscopic systems are of recent experimental
and theoretical interest because they give insight on the electronic
structure of the system in a noninvasive way, in contrast to the
transport studies\cite{tarucha} which are strongly dependent on the 
contacts necessary
to measure the electric current and the voltage.  Also, while the
transport properties of the quantum dots depend on the processes that 
occur at the Fermi level, the magnetic properties are determined by 
the whole energy spectrum of the mesoscopic system. The origin of the 
orbital magnetism consists in the permanent currents carried by the 
quantum eigenstates in the presence of the magnetic field and of the 
confinement. The currents are related to the energy spectra of the 
system under consideration.
For instance, the chirality of the current carried by one electron,
which gives the direction of the corresponding magnetic moment, can
be deduced from the dependence of the corresponding eigenenergy on the
magnetic field. In general, the spectral properties of the system
determine the sample-dependent characteristics of the magnetization.

The study of orbital magnetism under mesoscopic conditions was initiated
theoretically,\cite{vanrui,HSG} especially for predicting the role of
the boundaries of the samples.  An enhancement of the magnetization as
compared to the Landau diamagnetism of two-dimensional (2D) electrons
was demonstrated.\cite{HSG} More recently an exactly solvable model
of non-interacting electrons has been used for simulating persistent
currents and obtaining the orbital magnetization. This study has shown
that the currents are much more sensitive to the geometry of the system
than the associated magnetization.\cite{Inkson}   Bogacheck {\em et al.}
\cite{BSL} have shown, by analytical calculations, for circular and
non-circular quantum dots with many (noninteracting) electrons that
the oscillations of the magnetization with increasing the magnetic
field have a hierarchy of three characteristic frequencies, due to the
oscillations of the energy levels in the vicinity of the Fermi energy,
due to the Aharonov-Bohm interference, and due to the de Haas-van Alphen
(dHvA) effect, respectively.  The importance of the electron-electron
interaction in quantum dots being generally admitted, the consequences
for the magnetization of 2D electron systems have been studied for finite
and infinite modulated systems,\cite{Vidar1} and also for noncircular
dots.\cite{Vidar1,Ing}  In the quantum Hall regime, an enhancement
due to exchange and correlation was shown both experimentally and
theoretically.\cite{Heitman}  Other recent measurements of the dHvA
oscillations, showing clear sawtooth profiles, have been performed by
Wiegers {\em et al.} \cite{Wig} and by Harris {\em et al.}.\cite{Harris}
For quantum dots, an ingenuous indirect technique was used by Oosterkamp
{\em et al.},\cite{Oo} in order to evaluate the change in the magnetization
due to single electron tunneling, from transport measurements. Only
very recently the magnetization of arrays of quantum dots have been 
directly measured, but insufficiently understood.\cite{Schwarz}

Since the magnetic moment of an individual dot is extremely small,
the experimental endeavor is oriented nowadays towards the study of 
bigger ensembles in order to measure a cumulative effect.  In this case, 
the dots can be coupled to each other either only electrostatically, 
or also exchanging electrons by tunneling, so that, in principle, 
the magnetization of the ensemble of coupled dots may be very different 
from the scaled magnetization of an individual dot.  In order to be able
to distinguish between these two situations, one has to know beforehand
the behavior of one single dot and then to identify the coupling 
effects.

The aim of this paper is a parallel study of magnetic properties of 
single and double dots in the presence of the electrostatic coupling, 
resonant tunneling, and detuning. The double dot is viewed as a 
coherent quantum-mechanical system separated in two regions by a 
constriction that can be controlled by a parameter. The intra- and 
inter-dot electron-electron interaction is taken into account on the 
same footing, and the detuning between the two regions is realized by 
applying an external potential (or bias) on only one of them. The 
modifications in the energy spectrum, topology of local currents and 
the consequences for the orbital magnetization will be presented.

\section{The  tight-binding model and its range of validity}

We describe the 2D electron system in perpendicular magnetic field by a
discrete, tight binding (TB) Hamiltonian, defined on a rectangular 
lattice (or plaquette). The lattice consists of $N$ sites along 
the $x$-and $M$ sites along the $y$-direction, separated by an 
inter-site distance $a$, so that a lattice vector reads 
${\bf r}_{nm} =na{\bf e}_x+ma{\bf e}_y$, with $n,m$ integers. 
Choosing the symmetric gauge for the vector potential, the one-electron 
spinless Hamiltonian reads:
\begin{eqnarray}
&&H=\sum_{n,m} \left[~\epsilon_{nm} \vert n,m
\rangle\langle n,m\vert
+ t \left(e^{i\pi m\phi}\vert n,m\rangle\langle n+1,m\vert
+  e^{-i\pi n\phi}\vert n,m\rangle\langle n,m+1\vert\right) + h.c.~\right] \,.
\label{1}
\end{eqnarray}
In this expression, ${\vert n,m \rangle}$ is a set of orthonormal  
states localized at the sites ${\bf r}_{nm}$ and $\phi$ is the 
magnetic flux through the unit cell measured in quantum flux units, 
$\phi=Ba^2/\phi_{0}, \,\, \phi_{0} =1.43 \times 10^{-15}$ Weber. The 
hopping energy $t= \hbar^2/2ma^2$ will be considered as the energy unit 
and the lattice constant $a$ will be the length unit.  
The choice of the boundary conditions is essential for the spectral 
properties of the Hamiltonian (\ref{1}) and the orbital magnetism.  
The cyclic boundaries give rise to the Hofstadter
butterfly if the commensurability of the geometric and magnetic 
periods is ensured.\cite{Hofstadter}  For the finite system the natural 
boundary conditions are of Dirichlet type, in which case the 
commensurability condition is not necessary, and a quasi-Hofstadter 
spectrum, with edge states filling the gaps, is obtained.  Also the  
degeneracies specific to the usual Hofstadter spectrum (at $B\neq 0$) 
are lifted by the presence of the infinite walls.\cite{edgestates,AGMN}

Our discrete system can be easily tailored into various shapes, or 
into several subsystems, by removing some inter-site hopping terms, or 
by imposing infinite barriers at some sites. For instance, we can 
model one single quantum dot, and two or more coupled dots. One 
question is whether a real system can be reasonably described by such 
a discrete model. The interatomic distance for GaAs is 
$a\approx 0.5$\,nm which means that if we wish to reach the {\em atomic 
resolution} for a square dot of linear dimensions $L=100$ nm, we need 
a lattice of $1000 \times 1000$ sites.  Obviously this requires too 
much memory and computing power, and we must restrict to a smaller 
number of sites. In reality we need much less sites. For instance, a 
grid containing $20 \times 20$ sites corresponds to an inter-site 
distance $a\approx 5$ nm. In addition we have to specify the
strength of the magnetic field and the number of electrons. Obviously, 
the magnetic length $l_{B}=(\hbar /eB)^{1/2}$ must be larger than the 
inter-site distance~ $a$, or equivalently 
\begin{equation}
\phi < 1/2\pi \,.
\label{2}
\end{equation}
Therefore, to describe the square dot of 100\,nm $\times$ 100\,nm by 
a lattice of 20 $\times$ 20 sites, we have to consider $B < 20 $\,T, 
which actually covers a wide range of experimental interest.  As 
another example, a grid $10\times 10$  
satisfies the  same condition  only for smaller fields $B < 5$\,T.

Since the energy spectrum of the one-band 2D  tight-binding model 
contains  $N \times M$ (nondegenerate) eigenvalues, and is bounded in 
the interval $[-4t,4t]$, it can accommodate at most $N\times M$ 
spinless electrons. Thus, the condition (\ref{2}) must be supplemented
by the requirement $k_{Fermi} <1/a$ which
ensures that the cosine-type tight-binding spectrum approximates
well the parabola of the quasi-free electrons at low magnetic fields.
In terms of energies the condition can be written as
\begin{equation}
E_{Fermi} <2 t \,,
\end{equation}
the Fermi energy  being measured from the bottom of the spectrum. 
Evidently, a finer grid means a denser spectrum and a larger number of 
electrons. The second condition indicates  that a $20\times 20$
plaquette provides a reasonable approximation for a system of 
about 100 electrons.  In general we shall keep the 
number of electrons $N_e$ below $N\times M/3$.  Altogether, the 
above conditions show that the lattice model describes correctly a 
physical quantum dot if the electrons occupy only the states 
corresponding to the bottom-left corner of the 
quasi-Hofstadter spectrum.

Several papers used the tight-binding model to describe single quantum
dots as we do here,\cite{Canali,Berkov} or groups of coupled quantum
dots.\cite{Kotlyar,Kirczenow} (In the later case each dot was associated
to a single site of the lattice and had therefore no internal structure.)
The alternative to the discrete model is to consider a quantum dot defined
by a continuous confining potential, to expand the one-electron wave
functions in a set of basis functions and to diagonalize the corresponding
Hamiltonian matrix. But for the numerical calculations the basis has to
be truncated to a finite set, which is in fact equivalent to choosing a
finite number of sites in the tight-binding model. The bigger the basis,
the bigger the number of electrons afforded in the dot. The natural basis
functions are the Laguerre polynomials, and for reasonable results one
needs at least 2-4 times more basis functions than electrons. On the other
hand, the computational effort increases exponentially with the size of
basis set, such that for a non-circular and non-parabolic dot one can
hardly go beyond 4-6 electrons.\cite{Ing}  Instead, in the tight-binding
model we can consider 50-100 electrons or more, for any shape of the dot.

\section{Magnetization of a single dot}

In this section we discuss the relation of our tight-binding 
Hamiltonian (\ref{1}) and the orbital magnetization of the system.  
The one-body Hamiltonian can be formally written as 
\begin{equation}
H=\sum_{nm,n'm'} H_{nm,n'm'} |nm\rangle \langle n'm'| \,,
\end{equation}
where $H_{nm,n'm'}=H_{n'm',nm}^*$ are the matrix elements of $H$.  
We denote its eigenvalues by $E_{\alpha}$ and its eigenstates by 
$|\alpha\rangle$. The position operator for an electron on the lattice 
can also be written as
\begin{equation}
{\bf r} = \sum_{nm} {\bf r}_{nm} |nm\rangle\langle nm| \,,
\label{rop}
\end{equation}
and obviously ${\bf r}_{nm}$ and $|nm\rangle$ are eigenvalues and 
eigenvectors of the position operator.  
Then, the operator associated with the current carried by one electron 
of charge $e>0$ is 
\begin{equation}
{\bf J}=-e{\bf \dot r}=\frac{ie}{\hbar}[H,{\bf r}] 
=\frac{-ie}{\hbar} \sum_{nm,n'm'} H_{nm,n'm'}
({\bf r}_{nm}-{\bf r}_{n'm'}) |nm\rangle\langle n'm'| \,,
\label{jop}
\end{equation}
and obviously any eigenstate of $H$ has a zero average current,
$\langle\alpha|{\bf J}|\alpha\rangle = 0$.
The terms of the sum in Eq.\ (\ref{jop}) can be interpreted as the 
currents from the site $nm$ to the site $n'm'$, 
\begin{equation}
J_{nm,n'm'}=
\frac{-ie}{\hbar} \left[ \,  H_{nm,n'm'}
({\bf r}_{nm}-{\bf r}_{n'm'}) |nm\rangle\langle n'm'|
+  \, H_{n'm',nm}
({\bf r}_{n'm'}-{\bf r}_{nm}) |n'm'\rangle\langle nm|
\right] \,.
\label{jlop}
\end{equation}
With Eq.\ (\ref{jlop}) we can calculate the distribution of the 
current within the system, while Eq.\ ({\ref{jop}) is sufficient to 
define the magnetization operator for one electron, as the standard 
orbital magnetic moment, 
\begin{equation}
{\bf M} = \frac{1}{2} \, {\bf r} \times {\bf J} \equiv 
\frac{1}{2} \, (x J_y - y J_x) \, {\bf e}_z \,.
\end{equation}
For a given eigenstate $\alpha$ of the Hamiltonian the average 
magnetization is obtained, by combining Eqs.\ (\ref{rop}) and 
(\ref{jop}), as
\begin{equation}
\langle\alpha|M|\alpha\rangle=
-\frac{2e}{\hbar}\sum_{nm,n'm'} n m' \,
{\rm Im} \left[ H_{nm,n'm'} \langle\alpha|nm\rangle
\langle n'm'|\alpha\rangle\right] \,.
\label{malpha}
\end{equation}
This result does not depend on the way the sites are coupled 
(nearest neighbors, next-nearest neighbors, etc.), nor on the presence 
of the Coulomb interaction. It holds as long as we use a one-body 
Hamiltonian.  In particular, in the Hartree approximation, the 
interaction changes the states
$|\alpha\rangle$, but not the current operator Eq.\ (\ref{jop}), which
is insensitive to the diagonal matrix elements of the Hamiltonian.
(However, in the Hartree-Fock approximation the interaction becomes
visible also in the current operator.)

At fixed magnetic field, the ground state magnetization $M_g$
is calculated by summing up the  individual contributions
$M_{\alpha}$ of all occupied eigenstates:
\begin{equation}
M_g(B)= \sum_{E_{\alpha} \le E_F} M_{\alpha}.\\
\end{equation}

In the localized representation the current density {\bf j(r)}  reads:
\begin{equation}
{\bf j(r)}=\frac{-i e}{\hbar} \sum_{nm} \delta({\bf r} - 
{\bf r}_{nm})\sum_{n'm'}
H_{nm,n'm'} ({\bf r}-{\bf r}_{n'm'})|nm\rangle\langle n'm'| + h.c. 
\end{equation}
The TB model provides a quick proof of the fact that 
${\bf M}_{\alpha}$ defined above coincides with $\langle\alpha|~
(-dH/dB) ~|\alpha\rangle$, and thus, from the
Feynman-Hellman theorem, with $dE_{\alpha}/d\phi$. Indeed, 
from (\ref{1}):
\begin{equation}
{dH\over dB}= {a^2\over\phi_{0}} {dH\over d\phi} =
{i \pi a^2 t\over\phi_{0}} \sum_{nm}
[~m e^{i\pi m\phi} \vert nm\rangle\langle n+1,m\vert - 
  n  e^{-i\pi n\phi}\vert nm\rangle\langle n,m+1\vert~] + h.c.
\end{equation}
which gives immediately Eq.\ (\ref{malpha}). The same result can be 
obtained by defining the magnetization as the the magnetic-moment 
density $(1/2m_{\rm eff}){\bf r}\times{\bf j(r)}$ integrated over 
the whole area.\cite{DOT}

The sign of $M{_\alpha}$ is determined by the chirality of the
corresponding eigenstate, that is  by  the sign of the derivative
$dE_n/d\phi$, which can be easily observed from the quasi-Hofstadter
spectrum. The edge and bulk states coexist at strong magnetic fields,
have opposite chiralities, so that their contributions to the total
magnetic moment have different signs, and eventually may cancel each
other.  An example is given in Fig.\,1 which shows the  magnetization
carried by the first 120 eigenstates of a $20\times 20$ plaquette at
$\phi=0.1$, corresponding to a square quantum dot of width about 100 nm,
at $B\approx 15$\,T.

Since the bulk and the edge states generate intercalated bands and
quasi-gaps in the spectrum, one may guess an oscillatory behavior
of the total magnetization $M_g$ of the quantum dot as a function of
the number of electrons accommodated inside. For an infinite system
these oscillations become the well known sawtooth profile of the total
magnetization, the dHvA oscillations in the regime of strong magnetic
fields.  Recent experiments have shown them for extended, theoretically
infinite samples,\cite{Wig,Heitman} but, to our knowledge, nothing
has been reported yet for finite samples, i.\,e. when the contribution
from the boundaries of the system is expected to be important, except
theoretical results.\cite{BSL,Vidar1,NAZ}

In the low field regime the situation is changed: the spectrum of our
system is no longer organized into bands and gaps, and the consecutive
states may have alternating chiralities. Therefore the summations of
all contributions must give rise to much faster, eventually chaotic
oscillations of the magnetization as a function of the number of 
electrons. This case was studied by Shapiro, Hajdu and 
Gurevich\cite{HSG} by calculating the magnetic susceptibility in the 
grand-canonical ensemble. Their result - for square geometry - is a 
rapidly oscillating comb-type picture which demonstrates an 
enhancement compared to the Landau susceptibility.\cite{HSG}  Here  
we calculate $M_g$ as a function of $N_e$ at $\phi=0.0001$ 
($B\approx 0.015$\,T) and obtain also a comb-like
picture which is shown in Fig.\,2.  

From an experimental point of view, it is more interesting to analyze the
magnetization as a function of the magnetic field $B$. With increasing
$B$, but for a constant number of electrons, the magnetization also
oscillates changing the sign when the Fermi level crosses regions
of different chiralities.  The internal mechanism can be understood
from Figs.\ 3(a)-(b) which show the quasi-Hofstadter spectrum and the
magnetization of a $10 \times 10$ plaquette occupied by 10 electrons.
One notices that the spectrum contains numerous anti-crossing  points
where the slope of the energy levels change suddenly, together with the
chirality of the corresponding current.  In fact, at these transitions
the states change from edge to bulk states, or vice versa.  However,
for the total magnetization only the anti-crossings at the Fermi level
are important. The change of sign of the magnetic moment of that state
changes abruptly the total magnetization. Instead, the anti-crossings
below $E_F$ do not change the total magnetization because there is always
a compensation between two adjacent levels.  Indeed no compensation
occurs at the Fermi level.

For a large number of electrons, the Fermi level increases to a region
with a huge number of anti-crossings, yielding fine and superfine
oscillations of the magnetization obtained by Bogacheck {\em et al.} for
dots with 2500 electrons.\cite{BSL}  We approach this regime in Figs.\
4(c)-(d), for 100 electrons on a $20 \times 20$ plaquette.  The big
oscillations of the magnetization are precursors of the dHvA
effect occurring at strong magnetic fields, while the small
oscillations reflect the anticrossing points in the energy spectrum
at the Fermi energy.

Some disorder, if present in the system, would
lift the degeneracies but would also smoothen the anti-crossing regions.
Consequently, the oscillations of the magnetization would become also
smoother than shown in our Fig.\,3, and indeed, the same would happen
at a finite temperature.  We shall see in the next section that the
electron-electron interactions have a similar effect.

The analysis of the magnetization based on the spectral properties is
done here for a square geometry and a tight-binding model. The results
obtained in Ref.\ \cite{Ing} prove however that the conclusions are
model independent. In that paper, a similar behavior showing smooth
regions and jumps of the magnetization as a function of the field have
been found for circular and elliptic dots with 2-5 electrons, 
in the continuum model based
on the Darwin-Fock Hamiltonian. It was shown that the positions of the
jumps in magnetization can be identified also from the maxima of the
total energy as a function of magnetic field.

As function of $1/B$, the dHvA oscillations are almost
equidistant, with amplitude and period depending on the number of
electrons $N_e$, as shown in Fig.\ 4(a).  In the scaled variables $M_{g}/N_e$ and
$N_{e}/B$ we find that all the curves coincide rather well, which
is shown in Fig.\ 4(c).
From this scaling property one concludes that the period of the dHvA
oscillations is $\Delta(1/B)=\mbox{const}/ N_{e}$. This result differs from the
usual problem of dHvA effect in metals where the period is proportional
to $1/E_F$, while $E_F$ is considered  in good approximation independent
of $B$.  It turns out that for small confined systems  this is not true,
the Fermi energy exhibiting large oscillations with maxima coinciding 
with the peaks of the magnetization, as illustrated in Fig.\ 4(b).  

The scaling of the magnetization curves suggests that the 
magnetization per particle of a quantum dot can be written as    
\begin{equation}
{M_g\over N_e}= f \Big({N_{e}\over B}\Big) ,
\end{equation}
where $f(x)$ is an oscillating function in the high field regime and 
it is nearly constant at low fields. 

In principle the spin may play an important role in the magnetization.
However, in the present paper we neglect the spin.\cite{BSL,Creffield}
The orbital magnetization in GaAs is enhanced due to the effective mass
by 14.9, compared to the spin contribution.  The spin magnetization is in
general small at moderate or low magnetic fields, when the Zeeman energy
is much smaller than the cyclotron energy.  In a first approximation
the spin contribution is independent on the current distribution and
consists in relatively weak oscillations around zero, see e.\ g. 
Fig.\ 6 of Ref.\ \cite{Vidar1}

As mentioned in the Introduction we intended to identify possible effects
produced by the tunneling and electrostatic coupling between the dots.
Such effects are important because the magnetization is measurable only
for ensembles, and not for individual dots.\cite{Schwarz} Both types of
coupling have a considerable influence on the orbital part of the wave
functions, and consequently on the orbital magnetization.  This justifies
more attention to the orbital magnetization for multi-dot systems.

\section{Double dots: tunneling and electrostatic  effects}

We consider now two coupled quantum dots, and label them 1 and
2. The inter-dot tunneling, 
electron-electron interaction, and detuning introduce specific
aspects in the distribution of charge and persistent currents
which affects the orbital magnetization. The double dot is 
sketched in Fig.\,5, and it is considered as a unique
coherent quantum  system  described by the Hamiltonian
\begin{equation}
H=H_1(\phi)+H_2(\phi, V_g)+\tau H_{12}+ H_{el-el} \,, 
\end{equation}
where $H_1$ and $H_2$ correspond to the individual dots and are the 
same as in Eq.\ (\ref{1}). The two dots are coupled both by the 
tunneling term $H_{12}$, and by the electron-electron interaction 
$H_{el-el}$ which here is considered in the Hartree approximation.  
In addition we also consider a gate potential $V_g$ applied on the 
second dot, i.\ e.\ a detuning parameter, which yields an energy 
offset between the two subsystems, and consists of an extra diagonal 
term in $H_2$.  

The inter-dot resonant-tunneling term is
\begin{equation}
H_{12}= \sum_{n_1m_1,n_2m_2} c_{n_1m_1,n_2m_2}
\vert n_1m_1\rangle \langle n_2m_2\vert + h.c. \,,
\end{equation}
where the sites $(n_1m_1)$ and $(n_2m_2)$ belong to the first, and 
to the second dot, respectively. The sites $(n_1m_1)$ and $(n_2m_2)$ 
are connected or not if $c_{n_1m_1,n_2m_2}=1$ or 0.

For the Coulomb interaction we use the Hartree approximation.  The
exchange interaction might be important at high magnetic fields (even if
the spin is ignored), usually well above 1 Tesla.\cite{Vidar1,Heitman} As
we have checked, in the Hartree-Fock approximation the exchange effects
are not important for the tight-binding model, in the regime studied in
this paper.  In a recent paper Creffield {\em et al.} have studied a 
continuous square quantum dot with two electrons in a weak magnetic field 
by exact diagonalization, and found a good agreement with the tight-binding 
model in the Hartree approximation.\cite{Creffield} 

The Coulomb interaction, in the Hartree approximation, reads
\begin{equation}
  H_{el-el}= U_c \sum_{nm}~\sum_{n'm'(\neq nm)} {N_{n'm'}\over {\sqrt{(n-n')
^2+(m-m')^2}}}~ \vert nm\rangle \langle n'm'\vert \,,
\end{equation}
where $N_{nm}$ is the mean occupation number of the site $r_{nm}$ 
and has to be calculated self-consistently with the energy levels. 
The Coulomb  energy, in units of $t$, becomes $U_c=e^2/\kappa a t 
\approx 1$ where we have used  the dielectric 
constant $\kappa=12.4$ and the effective mass $m_{\rm eff}=0.067$ as for 
GaAs.  However, for $U_c=1$ it is technically difficult to obtain the 
convergence of our iterative numerical scheme. Therefore, being in fact 
interested in qualitative results, we use  in our calculation a lower 
value, $U_c=0.4$, which still produces a strong perturbation of the 
noninteracting states.
If we assume the physical dimension of each quantum dot to be 50\,nm, 
and we choose a rectangular plaquette of $10\times 20$ sites to model 
the double-dot system, meaning $a \approx 5$ nm, our $U_c$ corresponds
to a Coulomb energy of about 8 meV. 


We first consider no exchange of electrons between the dots, i.\ e.\
$\tau=0$, and we focus our attention on the effects induced by the
electron-electron interaction. In principle, this interaction may
play a role in the magnetic properties of the dots since it produces
rearrangements of the electric charge and of persistent currents. One
may distinguish between intra-dot and the inter-dot interaction, the
latter being the electrostatic coupling between dots.  We would like 
to find out which one is more important from the point of
view of the orbital magnetism.

It has already been observed that in confined systems the bulk states
are more sensitive to the Coulomb interaction than the edge states.
\cite{Vidar1,MAMN}  By calculating the Hartree spectrum, for a fixed 
number of electrons, we find that the energy distance between the bulk-type 
states, which almost form energy 'bands', increases, while the energy of 
the edge states remains almost fixed. Therefore the Coulomb interaction 
mixes the energies of bulk-type and edge-type states, and thus states 
with opposite chirality become intercalated also in the 'bands'. This 
can be seen in Fig.\,6, even in the absence of the inter-dot 
interaction. A supplementary repulsion is added when the inter-dot 
interaction is taken into account, and further changes occur in the 
spectrum. One also notices that the electron interaction contributes 
to the  repulsion of the levels at the anti-crossing points.

The magnetic moments $M_\alpha$ of the Hartree states are sensitive 
to these effects and incorporate both the broadening of the bands and 
intercalation of states with opposite chirality.  We intend to find 
out the role played by the electron-electron interaction, in general, 
and by the inter-dot coupling, in particular. Comparing the two curves 
of $M_g$ versus $B$ in Fig.\,6, we observe that the inter-dot coupling 
attenuates the jumps and rounds off the peaks. This happens mainly at 
low fields. The comparison of the magnetization shown in Fig.\,3, 
where $U_c=0$, and those shown in Fig.\,6 indicate that the major 
interaction effect in the magnetization comes from the inter-dot 
coupling. Also, for intermediate and high magnetic fields, the 
saw-tooth shifts slightly, as was already noticed in Ref. \cite{Ing} 
However, the magnetic susceptibility  $\chi =dM/dB \big|_{B=0}$ is 
not affected by the presence of the electrostatic coupling. 
We shall see below that this is not the case in the presence of 
tunnel coupling.

We now consider the role of the tunneling, which means $\tau\neq 0$.
To be within the tunneling regime, we choose $\tau=0.4t$. The tunneling
now lifts the two-fold degeneracy of the eigenstates of the double dot,
and doublets of states appear now in the spectrum.  On the other hand,
the eigenfunctions and the associated persistent currents penetrate the
constriction and are distributed spatially over the whole area of the
double dot. Under these circumstances, the distinction between  edge-
and  bulk-states, which  is applicable for single dots, becomes 
improper.
Mixed states may also occur as shown in Fig.\,7 (middle panel).

This means that the chirality of a state can no longer be guessed
from the localization of that state in the bulk or at the edge of the
sample. In the absence of the interaction the current associated with
the lowest-energy eigenstate of the double dot is distributed in the
middle of each dot, as shown in Fig.\,7 (top panel). 
But the Coulomb repulsion pushes the current distribution towards the 
edges, as shown in Fig.\,7 (bottom panel). In the 
mixed state $\#\,46$ the current shows, in different regions, either 
clockwise or anti-clockwise circulation so that the total chirality of 
the state can only be found by the explicit calculation of the 
corresponding magnetization $M_{46}$.

The numerical calculation of the total orbital magnetization, shown 
in Fig.\,8, indicates that its dependence on the magnetic field is 
affected by the resonant coupling only in the low-field domain where 
the sharp peaks are again smeared (for 10 electrons per dot this 
occurs below $B\sim2$\,T). Unlike the case of electrostatic coupling, 
the resonant tunneling modifies the magnetic susceptibility. Indeed, 
for the situation presented in Fig.\,8, 
the slope at $B=0$ remains negative but is smaller in magnitude
compared to the $\tau=0$ case.

\section{Redistribution of charge in detuned double dots}

When one of the dots has a different confinement than the other, or is 
subjected to a supplementary gate potential $V_g$, its individual energy
spectrum is {\it detuned} with respect to the spectrum of the other
dot. For the sake of definiteness we consider a negative gate potential,
such that the dot 2 in Fig.\,5 gains positive electrostatic energy,
$-eV_g>0$, being pushed energetically upwards. We assume each dot is
of linear dimension $L=100$\,nm, and is described by a $10 \times 10$ 
lattice.

The role of the detuning was emphasized in connection with the 
transport properties of double quantum dots.\cite{Klimeck} The 
differences in the energy levels of the two dots can block the 
transfer of electrons. The absence of the resonance condition implies 
the absence of doublets in the spectrum (case discussed in the
previous section) and also a strong divergence of permanent currents 
at the constriction between dots.

By varying continuously $V_g$, one allows the redistribution of the 
electrons between the two dots. Even for a weak inter-dot tunneling 
(i.\ e.\ $\tau\neq 0$, but small), the magnetization behaves in an 
interesting fashion. In order to understand the physical process, 
let us consider the case of nearly isolated dots. 
The spectra of the two dots are identical, but shifted by $eV_g$.  
Then by changing the gate potential, a series of resonances occur and 
at each resonance, one electron is transferred from the dot 2 to 
the dot I. The transfer may be accompanied by a change of 
chirality if, for instance, the electron moves from 
an edge state to a bulk state. Then, at the corresponding value of 
the gate potential, the orbital magnetization of the whole system has 
a jump, together with the occupation numbers of each individual dot.  
The jump of magnetization can be positive or negative, 
depending on the chiralities of the initial and final states. Both 
situations are visible in the top panel of Fig.\,9.

At a fixed magnetic field, the energy spectrum of the double dot 
contains a multitude of anti-crossing points between the levels of 
the dot I, the horizontal lines in the bottom panel of Fig.\,9, 
and of the detuned dot 2, the lines with slope about 1. Indeed, 
in Fig.\,9 one can see that the jumps of the magnetization, and of the 
numbers of electrons in each dot, occur simultaneously with the 
condition $E_{N}=E_{N+1}$, $E_N$ being the highest occupied level 
of the double dot.  The redistribution of the electrons means that 
the system undergoes a transition from the configuration ($N_{1},N_{2}$) to
($N_{1}+1,N_{2}-1$), where $N_1$ and $N_2$ are the numbers of
electrons in the first and the second dot respectively, and
$N_{1}+N_{2}=N$ is the total number of electrons in the system.
Then, the jump of the ground state magnetization
can be expressed in terms of the magnetic moments carried by the
eigenstates of the dots:
\begin {equation}
\Delta M_{g}= M_{g}(N_{1},N_{2})-M_{g}(N_{1}+1,N_{2}-1) =
M^{(1)}_{N_{1}+1}-M^{(2)}_{N_2}  \,,
\end {equation}
where $M^{(1)}_{N_{1}+1}$ and $M^{(2)}_{N_2}$ are the magnetic moments 
carried by the states $|N_{1}+1\rangle$ and $|N_{2}\rangle$, in the 
dots 1 and 2, respectively.

This situation actually occurs only for weak inter-dot coupling.  
Obviously, a stronger coupling spoils the quantization of the number 
of electrons and of the magnetization, that is the steps are less 
sharp and the plateaus less evident, as shown by the dashed lines in 
Fig.\,9.

As a function of the magnetic field, the spectrum of a double dot 
develops a {\it dual} aspect obtained by the mixing of the two 
Hofstadter spectra of individual dots with a relative shift of 
$eV_g$.  For instance, some eigenvalues originating in the first 
quasi-gap of one dot overlap with the second band corresponding
to the spectrum of the second dot. A similar situation occurs also 
for higher energies. The effect of the mixing on the total 
magnetization is shown in Fig.\,10 (b) where many secondary 
peaks can be noticed as compared to Fig.\,3.
The differences are thus produced by the numerous anti-crossing 
points in the spectrum of the double dot which appear at the 
intersection of the two detuned quasi-Hofstadter spectra.
In closing this section we mention that by including the Coulomb 
interaction we obtained similar results, without important qualitative 
differences.

\section{Conclusions}

Our general aim was to demonstrate the correlation between the spectral
properties and the orbital magnetization of a quantum dot, by considering
that the orbital magnetization is the sum of the individual orbital
magnetic moments of all eigenstates.  The electrostatic and tunneling
coupling of two quantum dots also bring new effects.  We have noticed that
the anti-crossing points in the spectrum have special significance because
of the sudden change in the chirality of the electronic orbits occurring
at these points, with corresponding sign changes of the magnetic moments.

We have shown that interesting conclusions can be deduced even only
from the field dependence of the Fermi energy.  This is because the
anti-crossing effect at the Fermi level cannot be compensated by the
opposite contribution of the adjacent level, which is empty. So, we have
put forward the argument for which the tunneling transport data can be
relevant for the magnetic properties.  Oosterkamp {\it et al.} \cite{Oo}
took advantage of this fact and performed an indirect measurement of the
changes in the magnetization under resonant conditions in double dots. In
the experiment, by sweeping the gate voltage $V_g$ at different magnetic
fields $B$, one identifies all pair values $(V_{g},B)$ when a peak in the
tunneling current occurs. Then, a change in magnetization should occur,
as we have discussed in the previous section.  In  Ref.\,\cite{Oo}
the change $\delta M$ is calculated  as being proportional to $\delta
V_{g}/\delta B$.  The sign of $\delta M$ can be positive or negative as
we have also found  in Fig.\ 9 (a). We  have stated also that the sign
is determined by the relative chiralities of  initial and final states.

The experiments \cite{Oo} were done in the weak coupling limit when the
mixing of the two spectra is negligible and, intuitively, one may think
in terms of alignment of energy levels. In our approach, this condition
is not compulsory since the double dot is treated as a coherent quantum
system. In the case of a larger coupling $\tau$  the anti-crossings,
the depletion of the detuned dot and the changes in magnetization become
more smooth.

The electron-electron interaction, mainly the inter-dot component,
affects the distribution of the persistent currents and attenuates the
oscillations of the orbital magnetization especially at low fields.
Nevertheless, according to our results the magnetic susceptibility
is not influenced by the electrostatic coupling.
A scaling behavior of the quantum dot magnetization as a function
of magnetic field at different number of electrons has been identified in the
dHvA regime.

\acknowledgements
{M. N. and V. M. were supported by NATO-PC TUBITAK Programme, and B. T. 
by TUBITAK, TUBA, NATO-SfP, and KOBRA-001, at Bilkent University, Ankara.  
The research was also financed by the CNCSIS and National Research
Programme, Romania, and by the Icelandic Natural Science Foundation,
and the University of Iceland Research Fund.} 


\begin{figure}
\epsfxsize 14cm
\epsffile{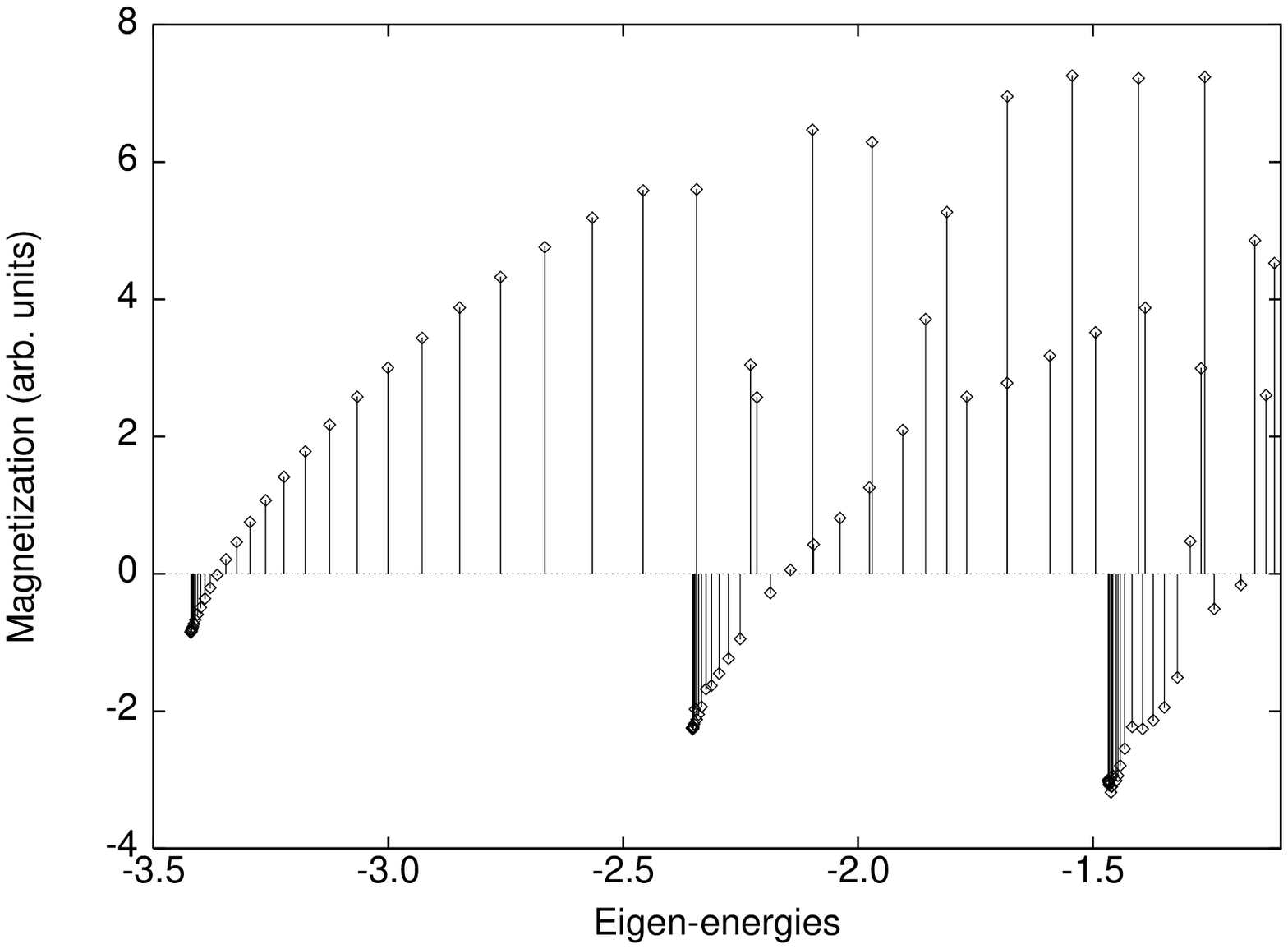}
\caption {Magnetizations $M_\alpha$ for each of the first 120 eigenstates,
for a lattice model of $20 \times 20$ sites, in the high-field regime,
$\phi=0.1$ (or $B \approx 15$ T, see text). The negative magnetizations 
grouped in bands correspond to bulk states, the positive ones fill the 
gaps and correspond to edge states.  }
\end{figure}

\begin{figure}
\epsfxsize 14cm
\epsffile{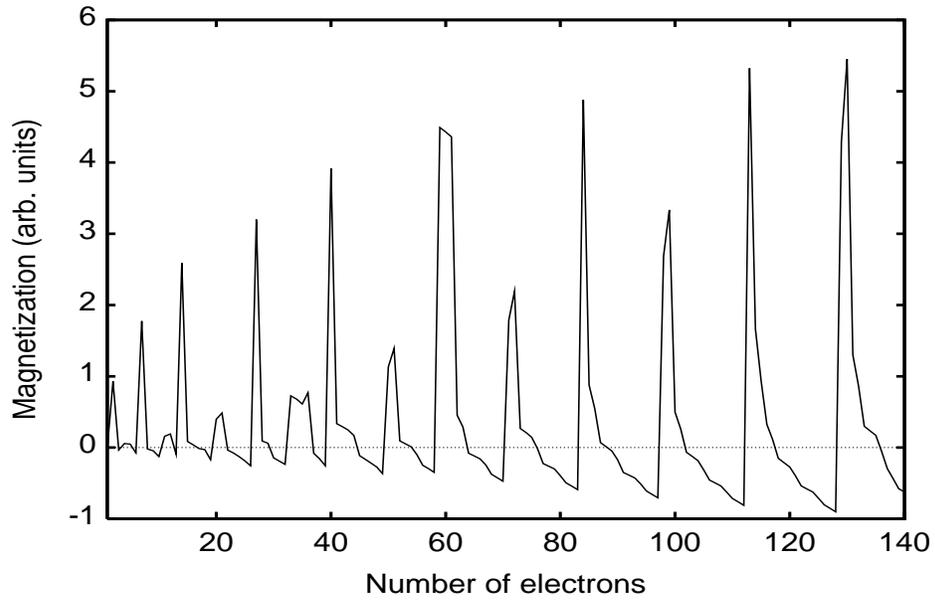}
\caption{ Total orbital magnetization as function of the number of
non-interacting electrons in the low field regime, $\phi=0.0001$
($B \approx 0.015$ T). The square dot of dimension $L=100$ nm is
represented in the tight-binding model by a plaquette  of
$20\times 20$ sites. } 
\end{figure}

\begin{figure}
\epsfxsize 17cm 
\epsffile{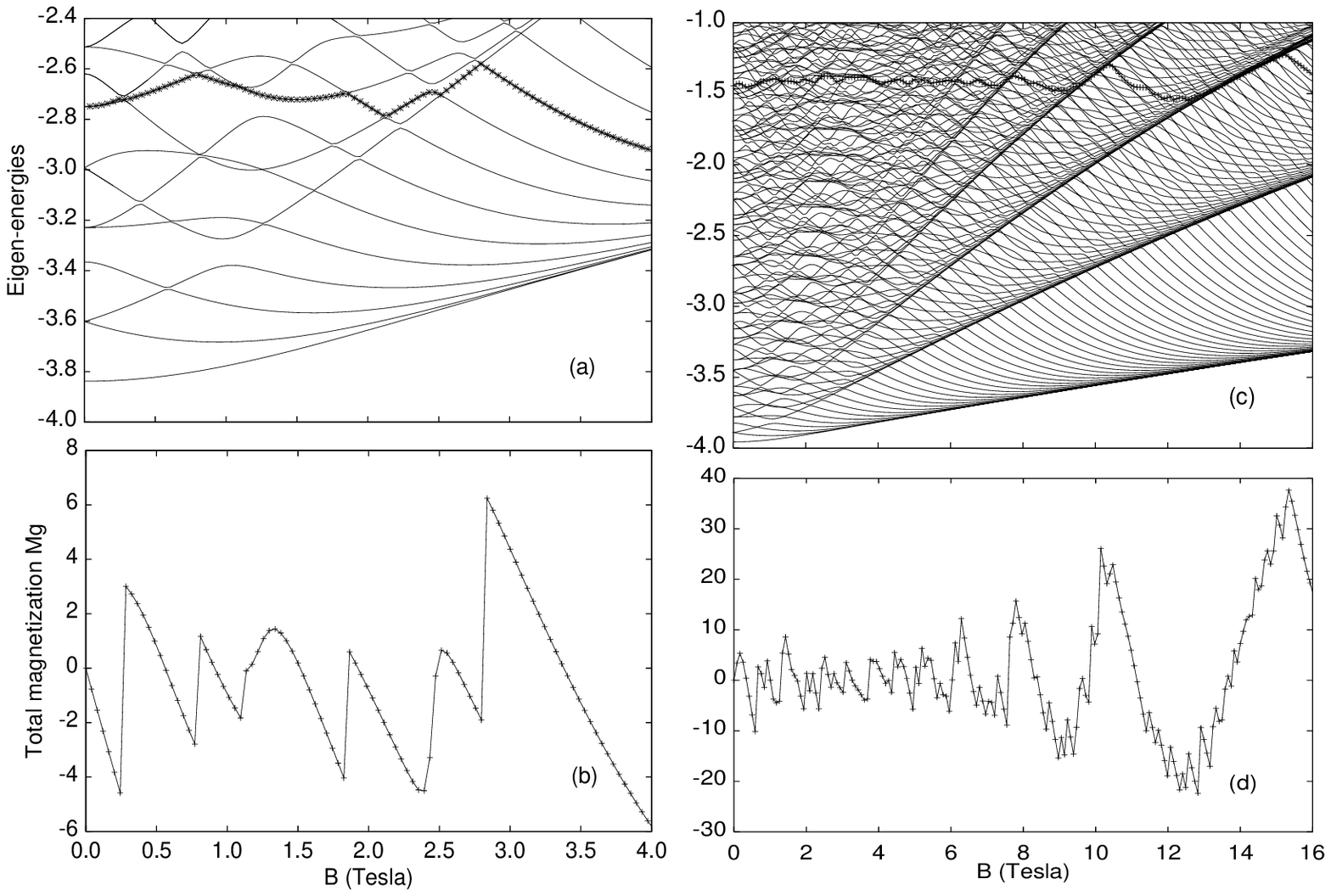}
\caption {(a) - The dependence of the spectrum and Fermi energy (with
crosses) on the magnetic field, for a square 10 $\times$ 10 plaquette 
with 10 non-interacting electrons. (b) - The corresponding total 
orbital magnetization.  The jumps in magnetization correspond to the 
change of the sign of $dE_{F}/dB$. (c) - The energy spectrum for a 
20 $\times$ 20 plaquette, with 100 non-interacting electrons, and 
(d) - the magnetization.}
\end{figure}

\begin{figure}
\epsfxsize 16 cm
\epsffile{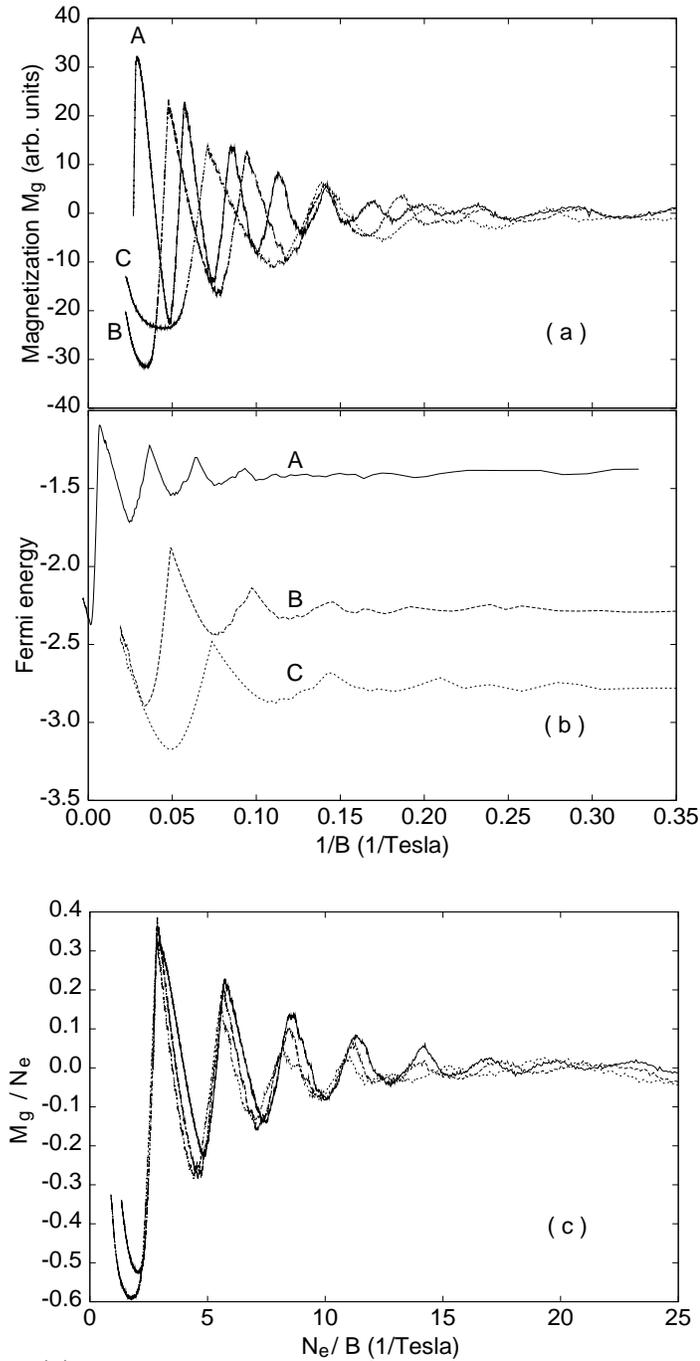}
\vspace{-2cm}
\caption 
{(a) - The magnetization of a square quantum dot $L=100$ nm versus the 
inverse magnetic field for different numbers of electrons:
100 (curve A), 60 (curve B), 40 (curve C).~
(b) - The Fermi energy for the same numbers of electrons.~
(c) - The scaled representation : magnetization per 
particle $Mg/Ne$ versus $Ne/B$;  the three curves  a,b,c
drops in a single one.}
\end{figure}

\begin{figure}
\epsfxsize 14cm
\epsffile{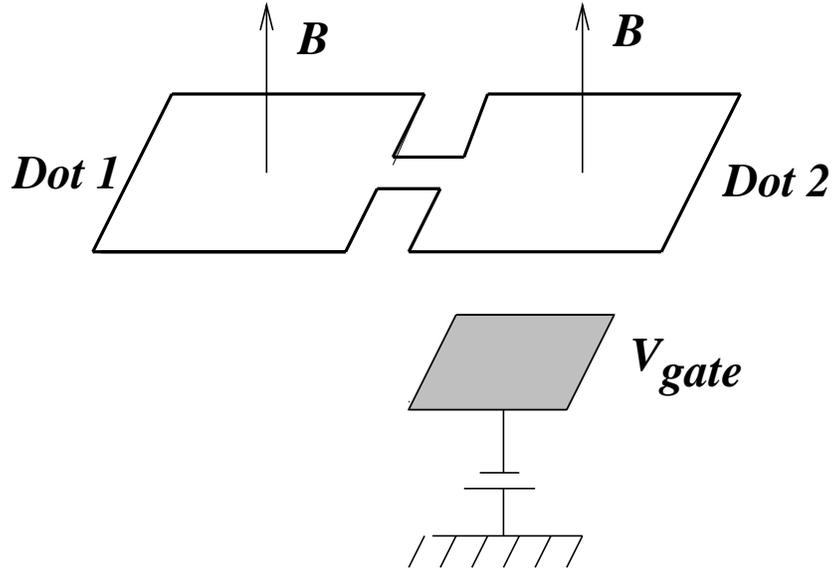}
\caption {Sketch of a double dot: tunneling may occur through the 
constriction (the central channel) and a gate voltage applied on the 
second dot  produces detuning effects.}
\end{figure}

\begin{figure}
\epsfxsize 16 cm
\epsffile{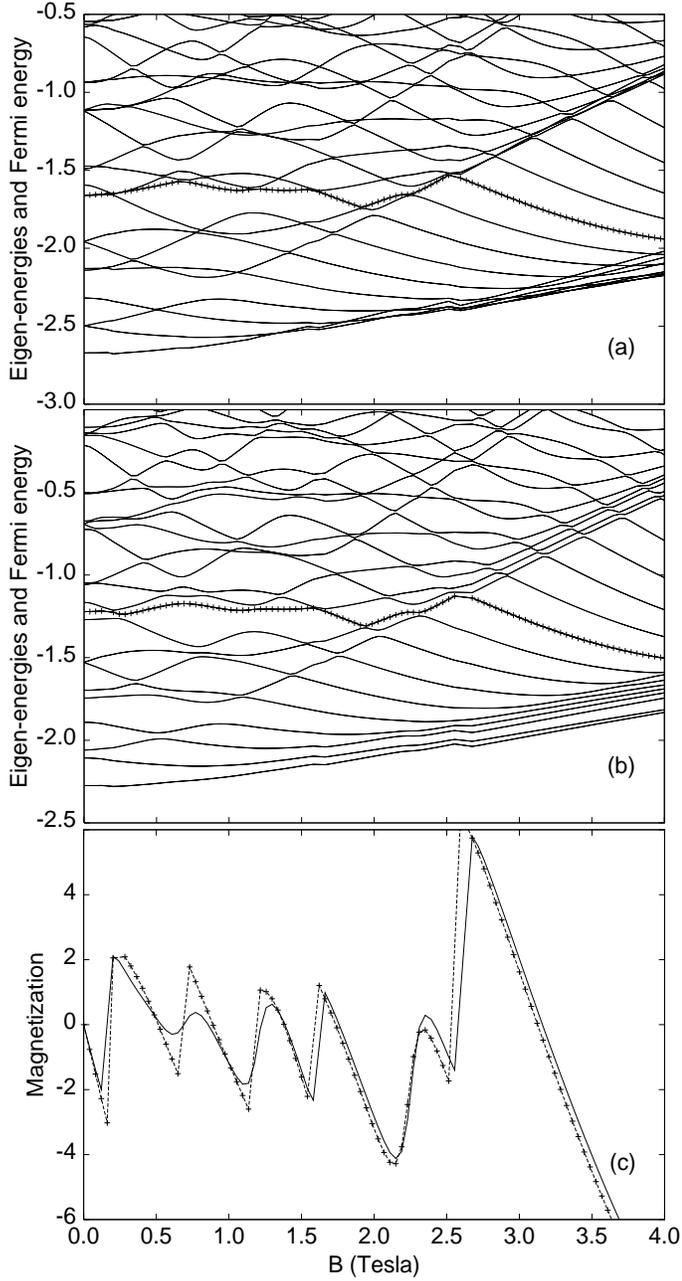}
\vspace{-3cm}
\caption {(a) - Energy spectrum, in the Hartree approximation,
in the presence of intra-dot electron-electron interaction, but with 
no inter-dot interaction.  The crosses show the Fermi energy.
(b) - The same, but with both intra- and inter-dot interaction. 
The dots are coupled only electrostatically, and the number of electrons 
is $N_e=10$ in each dot.  (c) - The magnetization for the two 
cases: for intra-dot interaction only, with the dashed line and crosses, 
and for the total interaction, with the solid line.
The whole system has $20 \times 10$ sites, and the interaction parameter
is $U_c=0.4$.
}
\end{figure}

\begin{figure}
\epsfxsize 16 cm
\epsffile{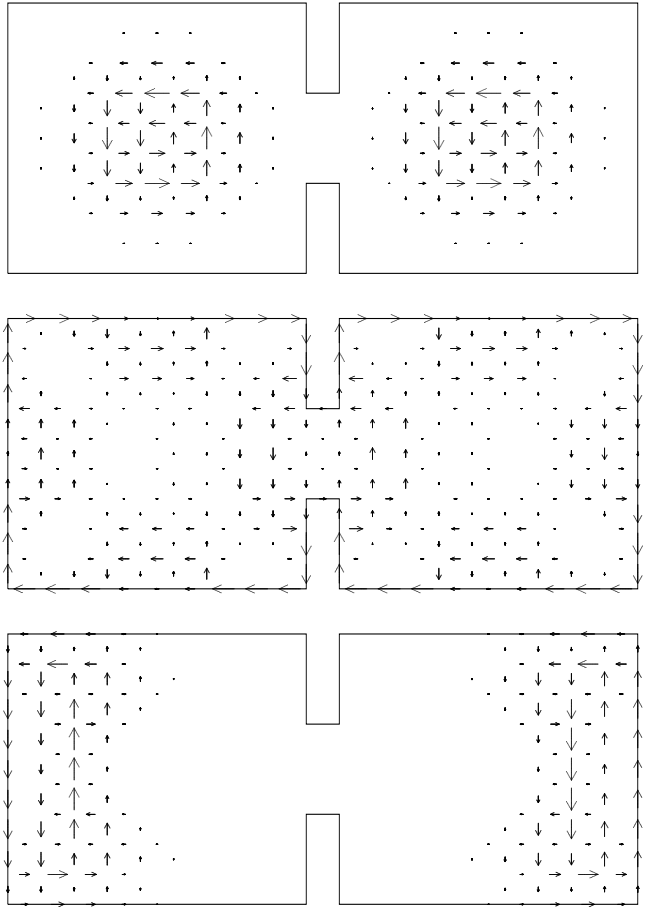}
\vspace{-5cm}
\caption {Persistent currents in double dot.
Top: The local current corresponding to  the eigenstate $\#\,1$ at 
$U_c=0$. Middle: The same for the eigenstate $\#\,46$. Bottom: The 
local current corresponding to the Hartree eigenstate $\#\,1$
at $U_c=0.4$}
\end{figure}

\begin{figure}
\epsfxsize 14 cm
\epsffile{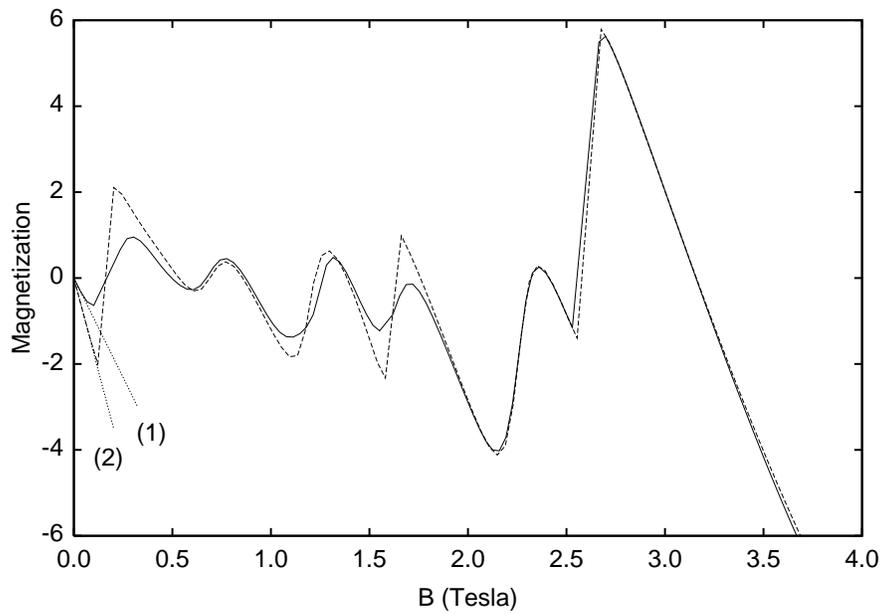}
\caption {The total magnetization of the double dot in the presence
of the electrostatic coupling : with tunnel coupling $\tau=0.4 t$ 
(full line) and without tunnel coupling $\tau=0$ (dashed). The 
straight lines (1) and (2) are the tangents at the magnetization 
curves in the low field domain for the two cases.}
\end{figure}

\begin{figure}
\epsfxsize 16 cm
\epsffile{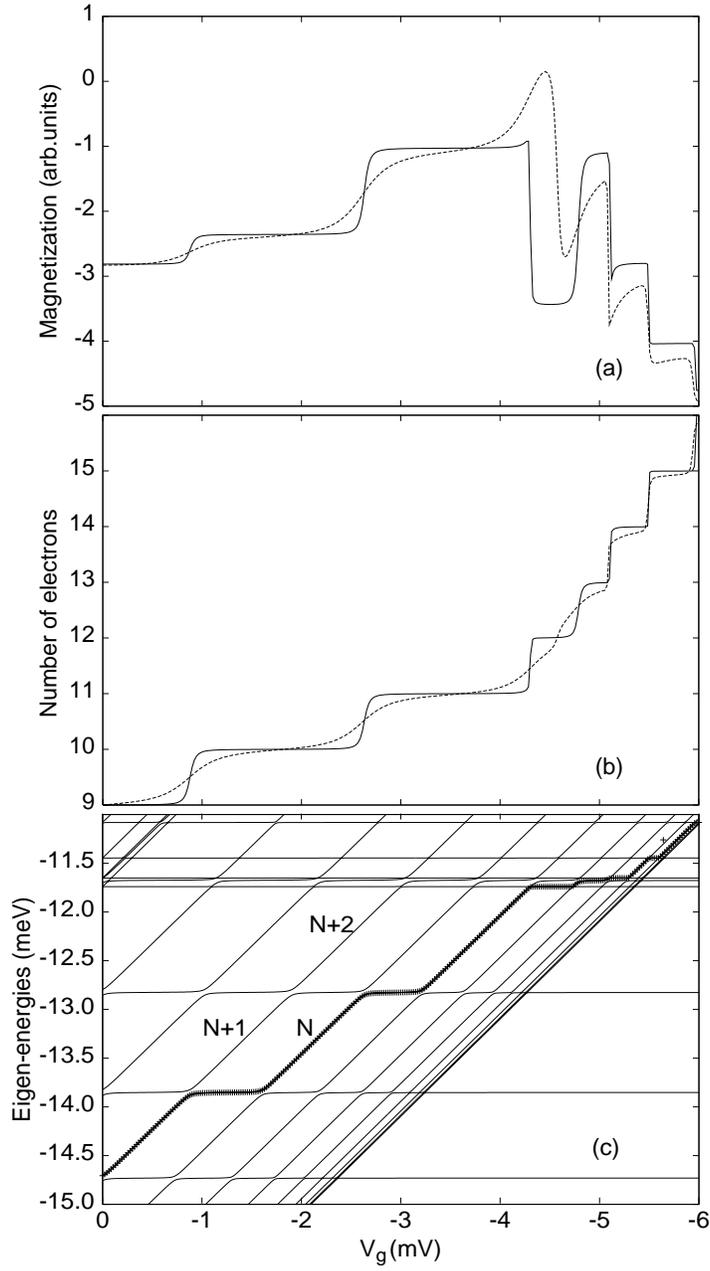}
\vspace{-2.0cm}
\caption {The magnetization (a), number of electrons in the 
nondetuned dot 1 (b) and a piece of the energy 
spectrum (c) versus the gate potential $V_g$. The following 
parameters have been used: $\phi =0.1$ (corresponding to 
$B \approx 4$\,T), $\tau= 0.1$ (the full line) and  $\tau= 0.4$
(the dashed line). The Fermi level corresponding to $N=18$ is also shown.}
\end{figure}

\begin{figure}
\epsfxsize 16 cm
\epsffile{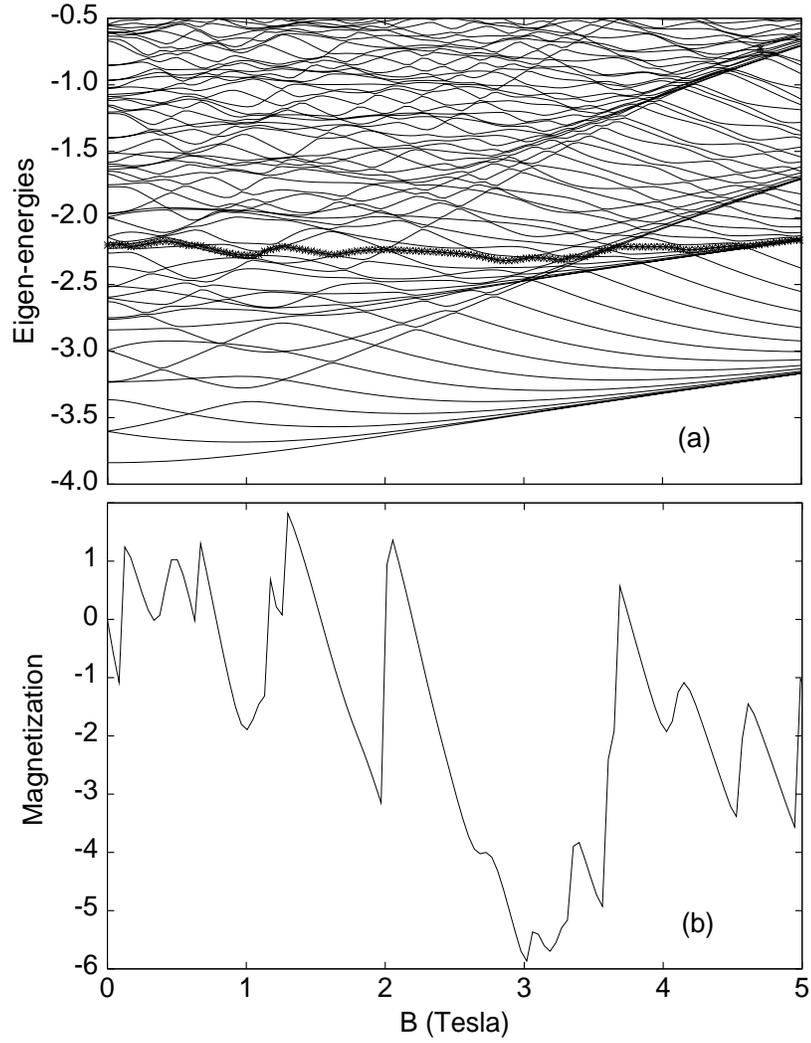}
\caption {The lower part of the energy spectrum (a) and the 
ground state magnetization (b) of a detuned double dot  
at $V_g =5$\,mV. The Fermi level corresponds to 20 electrons 
accommodated in the system. The tunnel coupling is $\tau= 0.4$.}
\end{figure}

\end{document}